\documentclass[12pt]{article}
\usepackage{graphicx}
\usepackage{amsmath}
\usepackage{amssymb}
\usepackage{latexsym}
\usepackage{hyperref}
\usepackage[sort]{cite}
\usepackage{bbm}

{\catcode`\|=\active 
  \gdef\Braket#1{\left<\mathcode`\|"8000\let|\bravert {#1}\right>}}
\def\bravert{\egroup\,\vrule\,\bgroup}
\newcommand{\be}{\begin{eqnarray}}
\newcommand{\ee}{\end{eqnarray}}
\newcommand{\bea}{\begin{eqnarray}}
\newcommand{\eea}{\end{eqnarray}}
\newcommand{\ben}{\begin{equation}}

\newcommand{\nn}{\nonumber}

\numberwithin{equation}{section}

\newsavebox{\ns}
\newsavebox{\dbrane}
\newsavebox{\dbshort}

\usepackage{epsfig}
\usepackage{amssymb}

\def\appendix{{\newpage\section*{Appendix}}\let\appendix\section%
        {\setcounter{section}{0}
        \gdef\thesection{\Alph{section}}}\section}

\newcommand\ba{\begin{eqnarray}}
\newcommand\ea{\end{eqnarray}}

\def\Dslash{\,\,{\raise.15ex\hbox{/}\mkern-12mu D}}
\def\Dbarslash{\,\,{\raise.15ex\hbox{/}\mkern-12mu {\bar D}}}
\def\delslash{\,\,{\raise.15ex\hbox{/}\mkern-9mu \partial}}
\def\delbarslash{\,\,{\raise.15ex\hbox{/}\mkern-9mu {\bar\partial}}}
\def\pslash{\,\,{\raise.15ex\hbox{/}\mkern-9mu p}}
\def\calDslash{\,\,{\raise.15ex\hbox{/}\mkern-12mu {\cal D}}}

\newcommand{\hh}{{1\over 2}}

\renewcommand{\ll}{_}
\newcommand{\uu}{^}
\newcommand{\pp}{\partial}
\renewcommand{\L}{\Lambda}
\renewcommand{\exp}[1]{{\rm exp}\left( #1 \right)}

\renewcommand{\d}{\delta}
\newcommand{\m}{\mu}

\renewcommand{\m}{\mu}
\newcommand{\n}{\nu}
\newcommand{\s}{\sigma}
\renewcommand{\t}{\tau}
\newcommand{\G}{\Gamma}

\renewcommand{\a}{\alpha}

\newcommand{\sqd}{^2}

\renewcommand{\hh}{{1\over 2}}

\newcommand{\eee}[1]{\ba{#1}\ea}
\renewcommand{\th}{\theta}
\renewcommand{\t}{\tau}

\renewcommand{\b}{\beta}

\newcommand{\apr}{{\alpha^\prime} {}}

\newcommand{\IZ}{\relax\ifmmode\mathchoice
{\hbox{\cmss Z\kern-.4em Z}}{\hbox{\cmss Z\kern-.4em Z}}
{\lower.9pt\hbox{\cmsss Z\kern-.4em Z}} {\lower1.2pt\hbox{\cmsss
Z\kern-.4em Z}}\else{\cmss Z\kern-.4em Z}\fi} \font\cmss=cmss10
\font\cmsss=cmss10 at 7pt
\newcommand{\inbar}{\,\vrule height1.5ex width.4pt depth0pt}
\newcommand{\IC}{{\relax\hbox{$\inbar\kern-.3em{\rm C}$}}}
\newcommand{\IQ}{{\relax\hbox{$\inbar\kern-.3em{\rm Q}$}}}
\newcommand{\IP}{\relax{\rm I\kern-.18em P}}
\newcommand{\Ione}{{\relax\hbox{$\inbar\kern-.39em{\rm 1}$}}}

\newcommand{\ed}{\dot{e}}

\renewcommand{\l}{\lambda}

\newcommand{\cc}{{\cal C}}

\newcommand{\ot}{\otimes}

\renewcommand{\cc}{{c_1}}

\newcommand{\lt}{\tilde{\lambda}}

\newcommand{\ct}{\tilde{c}}

\renewcommand{\cc}{c}

\renewcommand{\L}{\Lambda}

\newcommand{\cl}{{\cal L}}

\newcommand{\IR}{\relax{\rm I\kern-.18em R}}
\def\blfootnote{\xdef\@thefnmark{}\@footnotetext}

\renewcommand{\cc}[1]{\cite{#1}}

\newcommand{\bm}{\begin{matrix}}

\newcommand{\vws}{{V_{\rm ws}}}

\newcommand{\rr}[1]{(\ref{{#1}})}
\newcommand{\bbb}{\ba}
\renewcommand{\eee}{\ea}
\newcommand{\een}[1]{\label{#1}\ea}

\newcommand{\xxx}{(\xx)}

\newcommand{\forceheading}[1]{\ \\ \noindent {\bf #1} \nopagebreak \\  \nopagebreak }
\newcommand{\heading}[1]{\subsection{#1}}

\def\lrdd{\left(  }
\def\rrdd{ \right)}
\def\lsqq{\left[ }
\def\rsqq{ \right]}

\newcommand{\kket}[1]{\left | {#1} \right \rangle }

\def\bi{\begin{itemize}}
\def\ei{\end{itemize}}

\def\qb{\bar{q}}

\def\ed{\end{document}}

\def\cc{{\cal C}}

\renewcommand{\rr}[1]{(\ref{#1})}
\def\tb{\bar{\tau}}
\def\ct{{\cal T}}

\def\llt{\tilde{\l}}
\def\cc{\,}

\def\mfw{(-1)\uu{\rm F_{\rm w}}}

\def\xxx{\nn\\\nn\\}

\newcommand{\stopdoc}{\cheatsheet \end{document}}

\def\uhe{{UHE~}}

\def\susamt{{no~}}

\def\nota{}
\def\lmgg{\lrdd \IZ\ll 2\uu 5\rrdd\ll L }
\def\macm{{\bf m}}
\def\macn{{\bf n}}
\def\hhv{Ho\v rava}
\def\zzt{{\tilde{Z}}}
\newcommand{\yyt}[2]{\lrdd \zzt\uu{#1}{}\ll{#2} \rrdd}
\newcommand{\xxt}[2]{\lrdd Z\uu{#1}{}\ll{#2} \rrdd}
\def\ell{L}
\def\eff{F}
\def\ellnine{L{\rm 9}}
\def\effnine{F{\rm 9}}
\def\inine{I{\rm 9}}
\def\znine{Z9}

\def\HEnine{HE9}
\def\com{,\hskip-.6mm}
\begin{document}

\begin{titlepage}
\begin{flushright}
arXiv:0710.1628 [hep-th]
\end{flushright}
\vspace{15 mm}
\begin{center}
  {\Large \bf   A stable vacuum of the tachyonic $E_8$ string }  
\end{center}
\vspace{6 mm}
\begin{center}
{ Simeon Hellerman and Ian Swanson }\\
\vspace{6mm}
{\it School of Natural Sciences, Institute for Advanced Study\\
Princeton, NJ 08540, USA }
\end{center}
\vspace{6 mm}
\begin{center}
{\large Abstract}
\end{center}
\noindent
We consider tachyon condensation in unstable ten-dimensional heterotic string
theory with gauge group $E_8$.  In the background
of a lightlike linear dilaton rolling to weak coupling, we find
an exact solution in which the theory decays to a stable ground state.
The final state represents a new, modular-invariant perturbative
string theory, tachyon-free in nine spacetime dimensions with a spacelike dilaton gradient,
$E_8$ gauge group and no spacetime supersymmetry.  
\vspace{1cm}
\begin{flushleft}
October 9, 2007
\end{flushleft}
\end{titlepage}
\tableofcontents
\newpage

\section{Introduction}
Weakly coupled heterotic string backgrounds
with ten-dimensional Poincar\'e symmetry were classified some time ago 
\cite{tye,class1,class2,class3,class4,class5,class6,class7}.  Among them is the
unstable heterotic string with a single $E_8$ gauge group 
realized on the worldsheet as a level-two current algebra
(hereafter, the \uhe string).  
This background has been the subject of much interest.  
In contrast to the supersymmetric 10D heterotic string with
$E\ll 8 \times E\ll 8$ gauge group, the \uhe string has a single 
real tachyon $\ct$ and broken supersymmetry.

One prominent role for the \uhe string
emerges from the study of spacetime-destroying decay modes 
of nonsupersymmetric M-theory backgrounds
\cite{horfab}.   The {\it supersymmetric} heterotic background
has been interpreted as the $R\ll{11}\to 0$ limit 
of 11-dimensional M-theory compactified on an interval of
length $R\ll{11}$, on which the 11-dimensional gravitino has supersymmetric
boundary conditions.  Changing the boundary condition so that
the gravitino has half-integral rather than integral moding
on the interval produces a nonsupersymmetric theory in
10 dimensions, with a nonzero Casimir energy causing the
two endpoints of the interval to attract one another.

In \cite{horfab}, Fabinger and Ho\v rava proposed that the \uhe background
could be a limit of such a nonsupersymmetric configuration,
following a phase transition in which the $E\ll 8 \times \bar{E}\ll 8$
gauge group is broken spontaneously to a diagonal subgroup.  It was
also conjectured that the remaining singlet tachyon
can condense, generating
a ``decay to nothing,'' in the sense of
\cite{wittenbubble,Horowitz:2005vp,Hirano:2005sg,Headrick:2004hz,Gutperle:2002bp,Emparan:2001gm,Costa:2000nw,Aharony:2002cx}.
Utilizing and extending the methods of \cite{previous}, an exact solution
was studied recently by Ho\v rava and Keeler \cite{horkeel,horkeel2} describing such a decay
in the background of a lightlike linear dilaton rolling to weak 
coupling in the future.\footnote{The first exact 
heterotic bubble of nothing solution was written in \cite{previous2}.}

In this paper we follow the logic of 
\cite{previous,previous2,previous3,previous4,previous5} to analyze 
the condensation of the singlet tachyon of the \uhe theory.  
We show that, in addition to the solution studied in \cite{horkeel,horkeel2},
there exists another exact solution with the same initial state, leading to a 
notably different outcome.  Our solution is dimension-changing in the sense
described in \cite{previous2}, with a final state described by a stable string theory in
nine dimensions with no spacetime supersymmetry.

We analyze the \uhe theory in the background of a lightlike
linear dilaton rolling to weak coupling 
in the direction $X\uu - \equiv {1\over{\sqrt{2}}}(X\uu 0 - X\uu 1)$.
This theory admits deformations
with a tachyon $\ct$ growing exponentially
in the complementary lightcone direction $X\uu +\equiv {1\over{\sqrt{2}}}(X\uu 0 + X\uu 1)$.  
These deformations are exact solutions to the beta-function equations, 
and the motion of a string in this theory is integrable.
It has been suggested \cite{previous} that such null propagation of
the tachyon arises automatically
from the evolution of localized field inhomogeneities
in the background of a linear dilaton.

The organization of the paper is as follows.
After reviewing the \uhe theory and
the general properties of tachyon condensation in 
Section~\ref{review}, we consider two cases:  first, we review a solution in which
the tachyon $\ct$ depends only on the lightcone
direction $X\uu +$;  second, we go on to consider a more generic case, 
in which the tachyon profile is not assumed to preserve any
special symmetry.  
We review the more symmetric case (studied in \cite{horkeel,horkeel2}) in Section~\ref{case1}, 
where 8D Poincar\'e invariance is imposed on the directions 
$X_i$ transverse to the lightcone.  Under this restriction, the most generic 
solution to the equation of motion is of
the form $\Phi = -\frac{q}{\sqrt{2}} X\uu -,~ \ct = 
\m\, \exp{\b X\uu +}$, with $q\b = \sqrt{2} / \apr $, possibly including
fields that decay to zero at late times.

In Section~\ref{case2} we break Poincar\'e invariance
in the eight directions $X\ll i$ to
obtain late-time endpoints that differ qualitatively from the bubble of
nothing.
If the real tachyon $\ct$ is allowed to depend in an arbitrary way on $X\ll i$, 
$\ct$ generically develops zeroes along
loci of real codimension one in the eight directions spanned by $X\ll i$.  
To simplify the analysis, we take the
limit in which the field $\ct$ varies on long distance
scales in the $X\ll i$ directions, as compared to the string scale. 
In this limit, each component of the zero locus is well 
approximated by a flat, isolated component of codimension one, with the
tachyon $\ct$ varying linearly in the direction transverse to the component.
We show that the physics of our solution is that
of {\it dimension quenching} \cite{previous2}, where the number of spacetime
dimensions reduces dynamically from ten to nine.  In the limit
$X\uu + \to \infty$, strings propagate in only 8+1 dimensions,
confined to the locus $\ct = 0$ by a potential barrier that
becomes infinitely steep and high.

In Section~\ref{further}, we show that the stable, late-time limit is described by 
a novel perturbative heterotic string theory with $E\ll 8$
gauge symmetry, a linear dilaton gradient in a spacelike
direction, and \susamt spacetime supersymmetry in nine dimensions.
To simplify the exposition, we refer to this nine-dimensional heterotic 
theory as the \HEnine~theory.  
We calculate the one-loop partition function in the \HEnine~theory and
verify that it is indeed modular invariant. 
Unlike its 10D parent, the 9D heterotic theory supported inside the
bubble is {\it stable}.  It follows
that the dynamical dimension-reducing 
solution is indeed generic in the space of solutions: any normalizable on-shell perturbation
decays or disperses at late times. 
Section~\ref{conclusions} contains conclusions and general comments.

\section{Review of the unstable $E_8$ string in 10 dimensions}
\label{review}
In this section we review several properties of the \uhe string theory
that are salient to the present discussion.  We establish conventions, present the
worldsheet supersymmetry algebra and outline the free-fermion construction of the 
level-two $E_8$ current algebra.

\heading{Worldsheet dynamics of the UHE string}
The \uhe string \cite{tye,horfab}, like all perturbative
heterotic string theories,  is described in (super)conformal
gauge as a (0,1) superconformal field theory (SCFT).
There are ten embedding coordinates $X\uu\m$, transforming
in the standard way under 10D Poincar\'e invariance.  Each
embedding coordinate is paired with a right-moving worldsheet 
superpartner $\psi\uu\m$, and the total 
central charge of the right-moving SCFT is equal to $c_R = 15$. 

The left-moving side has an $E\ll 8$ current
algebra at level two, and a single Majorana-Weyl fermion $\lt$.
The current algebra has central charge $c_{\rm alg} = {{31}/ 2}$, and
the fermion $\lt$ contributes central charge $1/2$, so the total
central charge of the left-moving side, including 
the ten bosonic coordinates $X\uu \m$, is $c_L = 26$. 
The level-two $E\ll 8$ current algebra has a free-fermion representation
based on 31 left-moving Majorana-Weyl 
fermions $\lt\uu A$. 

The worldsheet degrees of freedom transform as follows under
the $(0,1)$ worldsheet supersymmetry:
\bbb
&&[Q,X\uu\m ] =  i \,\sqrt{{\apr}\over 2} \, \psi\uu \m  \ , 
\qquad 
\{Q,\psi\uu\m\} =  \sqrt{2\over{\apr}}\, \pp\ll + X\uu \m\ , 
\xxx
&&\{Q, \lt\nota \} = F\nota  \ , 
\qquad 
~~~~~~~~~~~ [Q,F\nota] =  i \,\pp\ll + \lt\nota \ , 
\xxx
&&\{Q, \lt\uu A \} = F\uu A  \ , 
\qquad 
~~~~~~~\, [Q,F\uu A] =  i \,\pp\ll + \lt\uu A 
\ .
\eee



\heading{Properties of the current algebra}
To better understand the discrete gauge symmetry 
of the \uhe string, we review the free-fermion
construction of the $E\ll 8$ current algebra at level two.
It was observed in \cite{tye} 
that there exists a discrete
symmetry ${\lmgg}$
(where the subscript $L$ indicates the group acting only on left-moving excitations)
acting on a set of 31 free fermions $\lt\uu A$, such that gauging ${\lmgg}$ leads
to a level-two current algebra with group $E\ll 8$.  With
${\lmgg}$ taken to be generated by $g\ll 1 ,\cdots, g\ll 5$, the 
action of ${\lmgg}$ on left-moving fermions is specified by:
\bbb
g\ll 1 = \s\uu 3 \ot 1 \ot 1 \ot 1 \ot 1 \ ,
&\qquad&
g\ll 2 = 1 \ot \s\uu 3\ot 1 \ot 1 \ot 1\ ,
\nn\\
g\ll 3 = 1 \ot 1 \ot \s\uu 3 \ot 1 \ot 1 \ ,
&\qquad&
g\ll 4 = 1 \ot 1 \ot 1 \ot \s\uu 3 \ot 1\ ,
\nn\\
 g\ll 5  = 1 \ot 1 \ot 1 \ot 1 \ot \s\uu 3  \ ,
&\qquad &
\eee
where these matrices act in
the basis $(\lt\nota,\lt\uu 1,\cdots,\lt\uu {31})$.
The single fermion $\lt\nota$ does not participate in the current algebra,
and is neutral under all elements of $\lmgg$.

The generators $g\ll i$ act with a minus sign on subsets of $(\lt\nota,\lt\uu A)$ in
blocks of 16 at a time.   As a result, every sector twisted
by an element of ${\lmgg}$ is level-matched.  As usual, we also gauge
the operation $\mfw$, which acts on all left- and right-moving fermions 
simultaneously with a $-1$.  The action of $\mfw$ defines an R-parity, meaning 
that it acts with a $-1$ on the right-moving supercurrent $G(\s\uu +)$.

There are no left-moving currents in the untwisted sectors, since any fermion bilinear 
(without derivatives)
is odd under at least some element in $\lmgg$.  All $E\ll 8$ currents come from the twisted
sectors.  Each of the 31 non-identity elements of $\lmgg$ defines a twisted NS sector in
which 16 of the 32 current algebra fermions are periodic.  Prior to imposing projections,
the number of fermion ground states is $2\uu{{{16}\over 2}} = 256$.  Imposing five independent
$g\ll i$ projections cuts the number of ground states by a factor of $2\uu 5$, leaving
eight ground states in each twisted sector, for a total of $8\cdot 31 = 248$.  The ground-state
weight in each of the twisted sectors is ${{16}\over{16}}$ from 16 periodic fermions, so there are
exactly 248 weight-one currents, generating the 248-dimensional Lie algebra $E\ll 8$.
The remaining fermion $\lt\nota$ is invariant under $\lmgg$ and transforms as 
an $E\ll 8$ singlet, but it plays a role as the left-moving matter 
part of the tachyon vertex operator.

\heading{One-loop partition function of the \uhe string}
We can summarize the spectrum of the \uhe string
by computing its one-loop partition
function.  We will take the standard notation
for the path integral on a torus of complex
structure $\t$, with two real fermions transforming
with signs of $(-1)\uu{a+1}$ and $(-1)\uu{b+1}$ around the spacelike and
(Euclidean) timelike cycles, respectively.  
The partition functions are:
\bbb
Z\uu a{}\ll b  \equiv {1\over{\eta(\t)}} \th\ll{ab}(0,\t) \ ,
\qquad
\zzt \uu a{}\ll b  \equiv {1\over{\eta(\tb)}} \th\ll{ab}(0,\tb) \ ,
\eee
for right- or left-movers, respectively.

In the untwisted sector -- that is, with no action of $\lmgg$ on fermions when
transported around the spacelike cycle -- the partition functions for the
left-moving fermions $(\lt\nota,\lt\uu A)$ are
\bbb
\ell\uu 0{}\ll 0 {}\uu{(\rm untw.)}(\tb) 
 &=& {1\over{32}}
\yyt 0 0 \uu 8 \lsqq \yyt 0 0 \uu 8
+ 31 \yyt 0 1\uu 8  \rsqq \ ,
\xxx
\ell\uu 0{}\ll 1 {}\uu{(\rm untw.)}(\tb) 
 &=& {1\over{32}}
\yyt 0 0\uu 8 \lsqq \yyt 0 0 \uu 8
+ 31 \yyt 0 1\uu 8  \rsqq  \ ,
\xxx
\ell\uu 1{}\ll 0 {}\uu{(\rm untw.)}(\tb) 
 &=& {1\over{32}} \yyt 1 0\uu{16} \ ,
\xxx
\ell\uu 1{}\ll 1 {}\uu{(\rm untw.)}(\tb) &=& 0 \ .
\eee
Here we have averaged over insertions of elements of $\lmgg$ to implement the
projection onto invariant states.
In the sectors twisted by a nontrivial element $g\in\lmgg$, we have
\bbb
\ell\uu 0{}\ll 0 {}\uu{(\rm twist=g)}(\tb)
 &=& {1\over{32}} \yyt 0 0 \uu 8 \yyt 1 0\uu 8 \ ,
\xxx
\ell\uu 0{}\ll 1 {}\uu{(\rm twist=g)}(\tb)
 &=& {1\over{32}} \yyt 0 1\uu 8 \yyt 1 0\uu 8 \ ,
\xxx
\ell\uu 1{}\ll 0 {}\uu{(\rm twist=g)}(\tb)
 &=& {1\over{32}} \yyt 1 0\uu 8 \lsqq   \yyt 0 0 \uu 8
+ \yyt 0 1\uu 8 \rsqq \ ,
\xxx
\ell\uu 1{}\ll 1 {}\uu{(\rm twist=g)}(\tb)
 &=& 0 \ .
\eee
Summing over twisted sectors, we obtain
\bbb
\ell\uu 0{}\ll 0 (\tb) &=& {1\over{32}} \yyt 0 0 \uu 8\lsqq \yyt 0 0 \uu 8 + 31 \yyt 0 1 \uu 8
+ 31 \yyt 1 0 \uu 8 \rsqq \ ,
\xxx
\ell\uu 0{}\ll 1 (\tb) &=& {1\over{32}} \yyt 0 1 \uu 8\lsqq \yyt 0 1 \uu 8 + 31 \yyt 0 0 \uu 8
+ 31 \yyt 1 0 \uu 8 \rsqq \ ,
\xxx
\ell\uu 1{}\ll 0  (\tb) &=& {1\over{32}} \yyt 1 0 \uu 8\lsqq \yyt 1 0 \uu 8 + 31 \yyt 0 0 \uu 8
+ 31 \yyt 0 1 \uu 8 \rsqq \ ,
\xxx
\ell \uu 1{}\ll 1  (\tb) &=& 0 \ .
\eee
As usual, the partition function for the right-moving fermions and superghosts is
$\xxt a b \uu 4$.
The full path integrals for the fermions and superghosts are then
\bbb
\eff\uu a {}\ll b  (\t,\tb) \equiv  \xxt a b  \uu 4 \ell\uu a{}\ll b \ ,
\eee
where we have summed over twisted sectors and projected onto $\lmgg$-invariant states.
Under $\tau \to \tau +1 $, the nonvanishing $\eff\uu a {}\ll b $ transform as
\bbb
\eff\uu 0 {}\ll 0 \to -\eff\uu 0 {}\ll 1 \ ,  
\qquad
\eff\uu 0 {}\ll 1 \to -\eff\uu 0 {}\ll 0 \ ,
\qquad
\eff\uu 1 {}\ll 0 \to +\eff\uu 1 {}\ll 0 \ ,
\een{ftforms}
while under $\tau \to -1/\tau$, the $\eff\uu a {}\ll b $ transform classically:
\bbb
\eff\uu a {}\ll b  \to \eff\uu b {}\ll a  \ .
\een{fclass}
The path integrals in the NS$\pm$ and R$\pm$ sectors are therefore
\bbb
I\ll{{\rm NS\pm}} \equiv \hh \lrdd \eff\uu 0{}\ll 0 \mp \eff\uu 0{}\ll 1 \rrdd \ ,
\eee
where the sign flip comes from the odd fermion number of the superghost ground state
in the NS sector.  The partition function in the Ramond sector is
\bbb
I\ll{{\rm R+}} = I\ll{{\rm R-}} = \hh \eff\uu 1{}\ll 0  \ .
\eee

The path integral over bosons and reparametrization ghosts 
is independent of the sector, and equal to
\bbb
i V\ll{10} (4\pi\sqd\apr\t\ll 2)\uu{-5} |\eta(\t)|\uu{-16} \ ,
\eee
for ten free embedding coordinates, where $V\ll{10}$ is the (infinite) volume of the
flat ten-dimensional spacetime.
Multiplying the two partition
functions
against the factor of ${{d\t d\tb}/{4\t\ll 2}}$ from the gauge-fixed
path integral measure \cite{joebook} 
and integrating over the modular fundamental region ${\cal F}$, we find
\bbb
\int\ll{\cal F} {{d\t d\tb}\over{16\pi\sqd\apr \t\ll 2\sqd}}
 (4\pi\sqd\apr\t\ll 2)\uu{-4} |\eta(\t)|\uu{-16}
  \lrdd I\ll {{\rm NS} +} - I\ll{{\rm R}\pm} \rrdd \ ,
\een{equation213}
with the minus sign in front of $I_{\rm R\pm}$ implementing the fermionic
spacetime statistics of the Ramond states.

The measure $d\tau \, d\bar\tau / \tau_2^2$ is itself modular invariant, 
as is the combination
$\tau_2 |\eta(\tau)|^4$.  The fermionic partition functions transform under
$\tau \to \tau +1$ as 
\bbb
I_{\rm NS\pm} \to \pm I_{\rm NS\pm} \ ,
\qquad 
I_{\rm R\pm} \to  + I_{\rm R\pm} \ .
\een{modinv1}
Since the $\eff ^a{}_ b$ transform classically under $\tau \to -1/\tau$, 
(\ref{fclass}) the combination 
\bbb
I\ll{\rm NS+} - I\ll{\rm R\pm} = \hh \lrdd \eff^ 0{}_ 0 - \eff^ 0{}_ 1 
			- \eff^ 1{}_ 0 \rrdd
\een{modinv2}
remains unchanged.  
It follows that the integral (\ref{equation213}) is modular invariant,
for either choice of GSO projection in the Ramond sectors. 

The factor $i V\ll{10} (4\pi\sqd\apr\t\ll 2)\uu{-5}$ in the 
modular integrand comes from the integral over bosonic zero modes:
i.e., the momentum integral for physical states.  Removing this and the measure
factor, and performing the $\t\ll 1$ integral that implements level matching, 
we obtain the partition function for the masses of physical states:
\bbb
Z\ll{\rm mass}(\t,\tb) &\equiv&
\sum\ll{{\rm physical}\atop{\rm states}} (-1)\uu{F_{\rm S}} \exp{- \pi \apr m\sqd \t\ll 2}  = 
\int\ll 0 \uu 1 d\t\ll 1  ~
|\eta(\t)|\uu{-16}
  \lrdd I\ll {{\rm NS} +} - I\ll{{\rm R}\pm} \rrdd 
\xxx
 &\equiv& Z\ll{\rm mass}\uu{\rm NS}(\t,\tb)  - Z\ll{\rm mass}\uu{\rm R}(\t,\tb) \ .
\eee
The expansion of these functions yields
\bbb
Z\ll{\rm mass}\uu{\rm NS}(\t) &=& (q\qb)\uu{-\hh} +  2\com 048   
	+ 148\com 752 ~(q\qb)\uu \hh
+ O\left(  (q\qb)\uu 1 \right) \ ,
\xxx
Z\ll{\rm mass}\uu{\rm R}(\t) &=& 3\com 968 + 9\com 404\com 416~
 (q\qb)\uu{1 } +  O\left(  (q\qb)\sqd \right)    \ ,
\eee
where we have defined $q\equiv \exp{2\pi i \t}$.

The first term in the NS partition function
corresponds to the tachyon $\ct$ in the spectrum, 
with a single state at mass $m\sqd = - {2\over{\apr}}$.
The second term corresponds to several different massless fields, including the
graviton $G\ll{\m\n}$, with $35$ polarizations, the B-field $B\ll{\m\n}$ with
$28$ polarizations, the dilaton with $1$ state, and the $E\ll 8$ gauge field
${\bf A}\ll \m$ with $8\cdot 248 =$ 1,984 massless states, for a total of 2,048 massless
states.  The next term represents the first massive level, with 148,752 states at
$m\sqd = {2\over{\apr}}$.

The leading term in the Ramond partition function represents a pair of Majorana-Weyl
fermions (one of each chirality), each in the adjoint of $E\ll 8$, for a total of
$2 \cdot 8 \cdot 248 =$ 3,968 physical states at $m\sqd = 0$.  The next term
appears at $m\sqd = {4\over{\apr}}$.  In particular, there is
no room for a gravitino, which, if present, would represent another $56$ physical 
states at $m\sqd = 0$.  


\section{The Ho\v rava-Keeler bubble of nothing}
\label{case1}
In this section we review the \hhv-Keeler exact bubble of nothing, with the intent of
contrasting it with the dimension-reducing solution described in the next section.

\heading{Linear dilaton background}
We consider the heterotic string in a lightlike
linear dilaton background, with dilaton gradient given by 
$V\ll{+} = V\ll{i} = 0$ 
and $V\ll{-} = - q / \sqrt{2}$.  We take $q$ to be positive, but its magnitude is frame dependent:
assuming it is positive, we can boost it to any
other positive value under a Lorentz transformation.

The worldsheet theory is conformally invariant, with
a free, massless Lagrangian given by
\bbb
{\cal L}\ll{\rm kin} = 
 {1\over{2\pi}} G\ll{\m\n}
\lsqq
{2\over{\apr}} 
 (\pp\ll + X\uu\m)
(\pp\ll - X\uu\n)
-  i \psi\uu\m (\pp\ll - \psi\uu\n) \rsqq 
-  {i\over{2\pi}}  \tilde{\l}\uu A (\pp
\ll + \tilde{\l}\uu A)
-  {i\over{2\pi}}  \tilde{\l} (\pp
\ll + \tilde{\l})
\ .
\een{hetfreelag}
The dilaton coupling takes the form
\bbb
{\cal L}\ll{\rm dilaton} =  \frac{1}{4\pi}
	\left( \Phi\ll 0 - \frac{q}{\sqrt{2}} X\uu -  \right) {\cal R}\uu{(2)}   \ ,
\eee
where ${\cal R}\uu{(2)}$ is the worldsheet Ricci scalar,
and $\Phi\ll 0$ is the value of the dilaton at $X\uu - = 0$.
To allow the closure of worldsheet supersymmetry off shell, we
supplement the action with kinetic terms for
the nondynamical auxiliary fields $F$ and $F\uu A$:
\bbb
{\cal L}\ll{\rm aux} = \frac{1}{2\pi}\left( {F\nota}^2 + ({F^A})^2 \right) \ .
\eee

\heading{Tachyon deformation}
To obtain the \hhv-Keeler solution,
we must deform the lightlike linear dilaton background
by letting the tachyon acquire a nonzero value 
obeying the equations of motion.  The single real tachyon
$\ct$ is a singlet of $E\ll 8$, and 
couples to the worldsheet as 
a superpotential
\bbb
W \equiv \llt \nota : \ct\nota (X) :\ ,
\eee
where the component action comes from integrating the
superpotential over a single Grassmann direction $\th\ll +$
(to form a $(0,1)$ superspace integral):
\be
{\Delta \cl} &\equiv& - {1\over{2\pi}} \int d\th\ll + ~ W
\nn \\ && \nn \\
&=& - {1\over{2 \pi}}
\lrdd F\nota :\ct\nota(X) :  - i \sqrt{\apr \over 2} \cc 
: \pp\ll{\m} \ct\nota (X):
\llt\nota  \psi\uu \m  \rrdd\ .
\ee
Integrating out $F\nota$ yields a potential of the form
\bbb
\Delta {\cal L} = - {1\over{8\pi}}  :\ct\nota(X) \sqd: \ ,
\eee
and a modified supersymmetry transformation 
$\{Q, \tilde{\l}\nota\} = F\nota = \hh :\ct\nota:$.

The linearized equation of motion for the tachyon is  
\bbb
\pp\uu \m\pp\ll\m \ct\nota - 2 V\uu\m \pp\ll\m \ct\nota
+ {2\over\apr} \ct\nota = 0\ .
\eee
The \hhv-Keeler solution takes the tachyon gradient to lie in
a lightlike direction:
\bbb
\ct\nota = \m\nota \, \exp{\b X\uu +}\ ,
\eee
with $q\b = \sqrt{2}/{\apr}$.  For diagrammatic reasons
(discussed in \cite{previous,previous2,previous3}), the worldsheet theory
defined by this superpotential has exactly vanishing beta function, so
the tachyon profile represents a solution beyond linearized order.

The bosonic potential on the string
worldsheet is positive-definite and increasing
exponentially toward the future:
\bbb
\Delta
{\cal L} = - {1\over{8\pi}} \m^2 \, \exp{2\b X\uu +}
+ {i\over{2\pi}} \sqrt{\apr\over 2} 
\b \m\nota \, \exp{\b X\uu +} ~\llt\nota \psi\uu + \ .
\eee
The classical worldsheet potential of this model is identical 
to that described in \cite{atnulltach1,atnulltach2,previous}.  
Consequently, classical solutions
for strings moving in the \uhe bubble of nothing are the same as
those in the bubble of nothing in Ref.~\cite{previous}.  That is, the solution
describes a Liouville wall moving at the speed of light in the negative 
$X\ll 1$ direction.  Every string state eventually meets the wall
and is accelerated outward to the left.

The worldsheet theory of the \uhe bubble of nothing differs 
from that of the bosonic bubble of nothing due to the presence
of worldsheet fermions. However, it is not straightforward to understand the physical meaning
of quantum effects contributed by the fermionic degrees of freedom.  Fermionic
interactions are only supported in the region of positive $X\uu +$, where string states are
energetically forbidden from penetrating.  In \cite{horkeel,horkeel2}, worldsheet techniques
were developed for studying the \uhe bubble of nothing deep inside the tachyon condensate.
In particular, the authors of \cite{horkeel,horkeel2} made
a non-standard gauge choice for the local worldsheet
supersymmetry in which the region of nonzero tachyon condensate could be studied with
greater ease.  The aim of \cite{horkeel2} was to understand the ``nothing phase''
of the \uhe string, possibly in terms of a conjectured topological phase of M-theory 
\cite{hortop} describing the region behind the \hhv-Witten wall \cite{horwit}.

Our focus is more pedestrian.  
The easiest observables to understand
are associated with regions whose future infinity has vanishing potential on
the string worldsheet.   In the next section, 
we will focus on solutions to the \uhe string that contain such
regions.   We study a different class of exact bubble solutions
to the \uhe string theory, where the interior of the bubble supports a
stable, nine-dimensional string theory rather than a topological or ``nothing'' phase.

\section{Dimension-changing solutions of the type UHE string}
\label{case2}
The bubble of nothing solution described above imposes 8D Poincar\'e symmetry
in the directions transverse to the lightcone defined by the dilaton
gradient $V_\mu$.  In this section we relax this assumption and examine 
solutions that break the 8D Poincar\'e symmetry.
Within this more general ansatz there are exact classical solutions of 
the \uhe model where
the total number of spacetime dimensions decreases from 9+1 to 8+1.
These solutions are qualitatively similar to the decays of the 
type HO${}^{+}$ background \cite{hellermanone} (i.e.,~the unstable supercritical 
heterotic model with nondiagonal GSO projection 
and gauge group $SO(32) \times SO(D-10) $).  
The similarity is that the tachyon is not
constrained to vanish at any particular locus defined in advance by the initial state.  
Our discussion of the transition from $9+1$ to $8+1$ dimensions
in the type UHE string applies equally well to the transition from $10+1$ to
$9+1$ dimensions in the type HO$^{+(1)}$ string described in \cite{hellermanone},
with the exception that in the present case the tachyon and dilaton gradients
are lightlike, rather than timelike.

Our initial conditions are such
that the $(9+1)$-dimensional theory at $X\uu 0 = -\infty$ has
FRW spatial slices in the form of the maximally
Poincar\' e-invariant nine-dimensional $\IR\uu  9$. 
The final theory will be a stable heterotic theory with gauge
group $E\ll 8$ (at level two) in
$8+1$ dimensions, with spacelike dilaton gradient and flat string-frame metric.
The solutions we study are exact to all orders in
$\apr$.  Furthermore, there is a readjustment of the dilaton and
the string-frame metric that comes from a one-loop effect on the
worldsheet, just as in the dimension-changing transitions discussed
in \cite{previous2}.

\subsection{Inhomogeneous profiles for the tachyon $\ct$}
At this point we allow the tachyon $\ct$
to vary in the eight dimensions $X\ll 2,\ldots , X\ll 9$ 
transverse to the lightcone directions $X\uu\pm$.  
We assume that the tachyon has a smooth vanishing locus
$\ct = 0$, and we take the limit in which
the typical scale of variation $|k\ll{2-9}|\uu{-1}$
of the tachyon is large compared to the string
length $\sqrt{\apr}$.
We can then approximate the tachyon as a linear function of the direction
normal to the locus $\ct = 0$.  Having done so, 
we can always perform an 8D Poincar\'e transformation
to set the zero locus precisely at $X\ll 9 = 0$.
We lose no generality, then, by taking 
\bbb
\ct\nota (X) = 
\sqrt{2\over\apr}\cc
\exp{\b X\uu +} \lsqq
\m X\ll 9 + O ( k X\uu 2 )
\rsqq\ ,
\eee
with
\bbb
q \b = \frac{\sqrt{2}}{\apr} \ .
\eee

Taking the long-wavelength limit and assuming a smooth vanishing
locus for the tachyon amounts to
dropping the $O (k X\sqd )$ terms.
The superpotential is then of the form
\bbb
W = \m\cc \sqrt{2\over\apr}  \cc
\exp{\b X\uu +} \, \lt \, X\ll 9    \ ,
\eee
which means that the interaction Lagrangian equals
\be
{\cal L}\ll{\rm int} &=& - {{\m\sqd}\over{4\pi\apr}} \cc\exp{2\b X\uu +}\cc 
: X\ll 9\sqd : 
+ {{i \m}\over{2\pi}} \exp{\b X\uu +} \cc
\tilde{\l}\nota
\lrdd  \psi\ll{9} + \b X\ll{9} \psi\uu + \rrdd\ .
\ee
This superpotential is the same as that discussed in the heterotic
section (Section 3) of \cite{previous2}, 
restricted to the case in which a single real tachyon acquires a vev.

The equations of motion take the form (with $i = 2,\cdots,8$):
\be
\pp\ll + \pp\ll - X\uu + &=& \pp\ll - \psi\uu + ~=~ 
\pp\ll + \pp\ll - X\ll i ~=~ \pp\ll - \psi\ll i 
~=~ \pp\ll + \tilde{\l} \uu A ~=~ 0\ ,
\nn\\ && \nn\\
\pp\ll + \pp\ll - X\ll 9 &=& - {{\m\sqd}\over 4} \cc\exp{2\b X\uu +}\cc X\ll 9 
			+  \frac{i \m \b \apr}{4} \exp{\b X^+} \lt \, \psi^+ \ , 
\nn\\ && \nn\\
\pp\ll + \pp\ll - X\uu - &=&  {{\m^2 \b}\over 4} \cc\exp{2\b X\uu +}\cc X\ll 9 \sqd
- {{i \m \b\apr }\over 4}  \cc\exp{\b X\uu +}\cc \tilde{\l}\nota \lrdd \psi\ll 9  + \b \cc
X\ll 9 \psi\uu + \rrdd\ ,
\nn\\ && \nn\\
\pp\ll + \tilde{\l}\nota &=&
{{\m}\over 2} \cc\exp{\b X\uu +}\cc
\lrdd \psi\ll 9 + \b X\ll 9\psi\uu + \rrdd \ ,
\nn\\ && \nn\\
\pp\ll - \psi\uu - &=& {{\b \m}\over 2} \exp{\b X\uu +} \cc
X\ll 9 \tilde{\l}\nota \ .
\eee
This 2D worldsheet theory is integrable at both the classical and 
quantum levels.  As a result, the properties of string trajectories 
in this theory are particularly simple, and can be summarized
as follows:
\bi
\item{In infinite worldsheet volume, the general classical
solution can be written in closed form.}
\item{In finite worldsheet volume, trajectories with
$X\uu \pm,~\psi\uu\pm$ independent of the spatial
worldsheet coordinate $\s\uu 1$ can be written in a simple closed form
involving Bessel functions (see, e.g., 
Eqn.~(2.17) of Ref.~\cite{previous2}).  The Virasoro constraints and
null-state gauge equivalences are sufficient to put a general
physical solution into such a form.}
\item{The classical and quantum
behavior of these solutions can be understood
from general principles, using the adiabatic and virial 
theorems.  The excitation
numbers in the massive modes $X\ll 9, ~\psi\ll 9, ~\tilde{\l}\nota$
are frozen into constant values at late times.}
\item{String behavior relative to the bubble wall
at $X\uu +\sim 0$ depends on asymptotic occupation
numbers.  Strings can enter the bubble wall and exist in the $X\uu + \to \infty$ 
phase if the occupation numbers in the massive modes are all equal to zero.  Otherwise, the string
is pushed outward along the bubble wall at $X\uu 0 \sim - X\uu 1$,
accelerated to the speed of light.}
\ei

\subsection{Effective worldsheet theory at large $X^+$}
As with the examples studied in \cite{previous,previous2,previous3,previous4,previous5},
the properties that render the theory exactly
solvable at the classical level also make the quantum theory particularly simple.
Namely, all connected correlators of free fields have quantum perturbation
expansions that terminate at one-loop order.
The structure of the quantum corrections can be summarized as follows: 
\bi
\item 
The two-point function for the light-cone fields $X\uu\pm$ and their superpartners $\psi\uu\pm$
are proportional to $G\uu{\m\n}$, so the propagators 
only connect ``$+$'' fields to ``$-$'' fields.
Propagators for the massless $X\uu\pm$ multiplets are therefore {\it oriented,} 
and we represent them as dashed lines with arrows pointing from $+$ to $-$.
The massive multiplet $X\ll 9,~ \psi\ll 9 ,~\lt\nota$ is correlated with itself,
so its propagators are represented by solid, unoriented lines.
\item
Fundamental vertices representing the
classical potential and Yukawa couplings
have arbitrary numbers of
outgoing dashed lines emanating from them, and exactly two solid lines. 
\item 
Every connected tree diagram with multiple vertices
therefore has the structure of an ordered
sequence of vertices with a single solid line passing
through, and arbitrary numbers of dashed lines
emanating from each vertex. 
\item 
Interaction vertices have only outgoing (dashed) lines, and no 
two vertices can be connected with a dashed line.  
A connected Feynman diagram can have either zero or one loop, but not two loops or more.  
\item
Connected loop diagrams consist strictly of a closed solid line
with dashed lines emanating from an arbitrary 
number of points.  This exhausts the set of
connected Feynman diagrams in the theory, and every connected correlator is exact at
one-loop order.
\ei

\heading{Dynamical readjustment of the metric and dilaton gradient}
Now we would like to study the dynamics of states that penetrate into the far interior of
the bubble, at $X\uu + \to \infty$.  These strings necessarily have all massive modes in their
ground states, frozen out with exponentially increasing mass.  
It is possible to integrate out $X\ll 9, ~\psi\ll 9$ and $\lt\nota$
exactly to obtain an effective action for the remaining worldsheet
degrees of freedom.  All but a finite number of terms have
canonical dimension greater than 2, coming from derivatives and fermions.  By scale invariance, 
such terms always appear dressed with real exponentials of $X\uu +$, with negative exponent.
We are ultimately interested in the limit $X\uu + \to  \infty$, so
all terms in the action of canonical dimension greater than two can be ignored.
Furthermore, no terms of canonical dimension less than two are generated, since
there are no such operators that are invariant under the $\lmgg$ symmetry 
and $(0,1)$ supersymmetry.

\begin{figure}[htb]
\begin{center}
\includegraphics[width=3.0in,height=1.5in,angle=0]{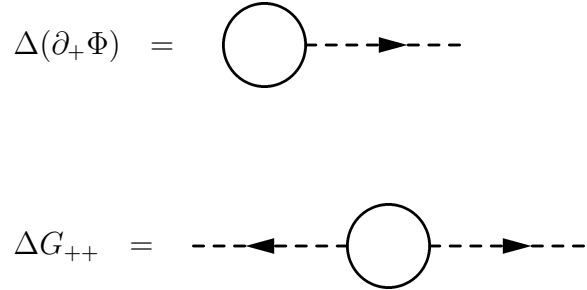}
\caption{Diagrams contributing to the nonvanishing 
renormalizations of the dilaton and metric, 
$\Delta(\partial_+ \Phi)$ and $\Delta G\ll{++}$.  Solid lines indicate massive fields, 
while dashed oriented lines represent propagators of the massless lightcone fields $X^\pm$.}
\label{renorm}
\end{center}
\end{figure}

The only terms generated that survive the $X\uu +\to\infty$
limit are thus a renormalization of the Ricci term $\sqrt{g} {{\cal R}^{(2)}}$, and a
renormalization of the kinetic term for $X\uu +,~\psi\uu +$.  In the
language of spacetime physics, these represent dynamical readjustments of the
string-frame metric and the dilaton due to the backreaction of the
condensing tachyon.  As described in \cite{previous2}, 
these readjustments arise entirely from one-loop renormalizations on the string
worldsheet, so they are exactly calculable.   The relevant diagrams are
depicted in Fig.~\ref{renorm}, where solid lines are drawn to indicate
massive fields, and dashed oriented lines indicate the massless $X^\pm$ fields.  
We will refer to quantities in the final-state theory with a hatted
notation.  The readjusted metric and dilaton are thus denoted by $\hat{G}$ and 
$\hat{\Phi}$, respectively, with the gradient of the readjusted dilaton denoted by
$\hat{V}_\mu$.

The calculation of the renormalized worldsheet couplings is
straightforward (see Refs.~\cite{hellermantwo,previous2} for details).
Defining $M\equiv \m \, \exp{\b X\uu +}$, the dilaton readjustment is
\bbb
\Delta \Phi =  {1\over 4} \ln \lrdd {M\over{\tilde{\m}}} \rrdd\ ,
\een{dilrenormone}
where $\tilde{\m}$ is an arbitrary mass scale entering the definition of the
path integral measure.
The renormalization can be expressed as
\bbb
\Delta \Phi = \lrdd {\rm const.}  \rrdd + 
{\b\over {4}}  X\uu +  \ ,  \qquad \Delta V_+ = \frac{\b}{4}  \ .
\een{dilrenormtwo}
As noted, there is also a nonzero
renormalization of the string-frame metric.
For a generalized mass term ${1\over{4\pi\apr}}  M (X)^2 ~X\ll 9\sqd$, where 
$M(X)$ depends arbitrarily on all coordinates other
than $X\ll 9$, the metric $G\ll{\m\n}$ is renormalized by an amount
\bbb
\Delta G\ll{\m\n} = {{\apr}\over 4} {{\pp\ll{\m}M  \pp\ll\n M}
\over{M\sqd}} \ . 
\een{metrenormone}
For $M = \m \cc\exp{\b X\uu +}$, this gives the renormalized metric
\bbb
\hat G\ll{++} = -\hat G\uu{--}  = {{\apr\b\sqd }\over 4}\ ,
\een{metrenormtwo}
with all other components unrenormalized.

The linear dilaton central charge at $X\uu + \to  \infty$ is therefore
given by (including the renormalized dilaton gradient $\hat V\ll\m$):
\bbb
c\uu{\rm dilaton} = 6\apr \hat{G}\uu{\m\n}
\hat{V}\ll\m \hat{V}\ll\n 
=  {{3  q\b\apr}\over{\sqrt{2}}}
- {{3  \b\sqd q\sqd \apr\sqd}\over 4}\ .
\eee
With $q\b = \sqrt{2} / \apr$, the final dilaton contribution to the central charge is
\bbb
c\uu{\rm dilaton} = {3\over 2} \ .
\een{finaldilhet}
We therefore find that exactly ${{3}/ 2}$ units
of central charge have been transferred from the
fields $X\ll 9,~ \psi\ll 9 , ~\lt\nota $
into the strength of the dilaton gradient at $X\uu + = \infty$.
The total central charge, including free-field and dilaton contributions,
is again the same at $X\uu + =  \infty$ as at $X\uu + = - \infty$, and in particular
equal to $(26,15)$.

\subsection{Stability properties of the tachyon solution}
\label{genericity}
We will demonstrate explicitly 
in Section~\ref{further} that the nine-dimensional string theory describing
the final state of the transition is {\it stable,} meaning it has no spatially
normalizable perturbations that grow with time.  
For the purposes of the discussion in this subsection, however, 
we will assume the stability of the final state.
No normalizable perturbation
of the transition as a whole should be able to affect the qualitative nature of the
endpoint: a perturbation of the full, time-dependent background will always evolve into
a perturbation of the late-time background.  The latter is stable, so the late-time
limit of such a perturbation cannot lead to a further instability changing the
phase of the final state.  In Ref.~\cite{previous4}, transitions having
this property were referred to as {\it stable transitions},  
and many of the transitions studied in \cite{previous2} are in fact stable.  In particular,
the transitions whose final states are spacetime-supersymmetric or two-dimensional
are such that a generic perturbation (satisfying the appropriate normalizability condition)
of the full, time-dependent solution will not
alter the solution qualitatively at late times.   
In this subsection we wish to refine this classification
by drawing a further important distinction between 
different types of stable transitions.  For purposes of our discussion, we will consider
bubble of nothing solutions to be stable transitions, in that no normalizable perturbation
changes their qualitative behavior.

\begin{figure}[htb]
\begin{center}
\includegraphics[width=3.2in,height=2.5in,angle=0]{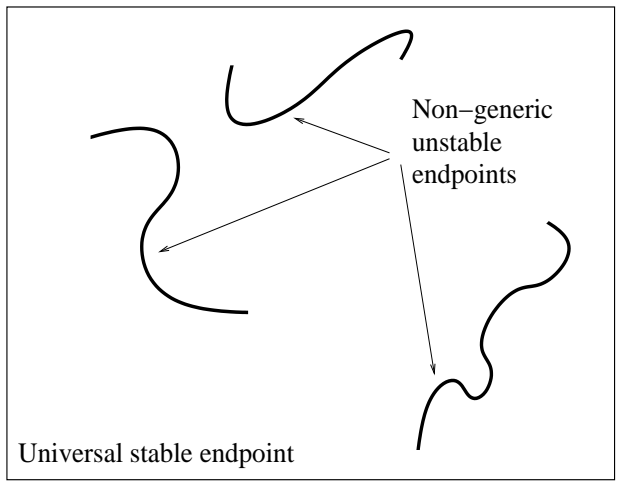}
\caption{Schematic phase diagram of endpoints in absolute-stable transitions. 
Contours represent loci of non-generic unstable endpoints of the transition, 
which are always fine-tuned.  Untuned transitions land in a universal stable theory,
away from the loci of unstable endpoints.}
\label{absolute}
\end{center}
\end{figure}

\forceheading{Absolute vs. local stability}
We first define {\it absolutely stable} transitions  
as those that exhibit a stable
final endpoint whose qualitative nature is 
completely determined in advance by the nature of the initial state.
For instance, the transitions from supercritical type HO$^{+/}$
(which has diagonal GSO projection) on certain orbifolds
to supersymmetric type HO string theory in ten dimensions have the property of absolute
stability.  Even a large, non-infinitesimal change in the initial configuration of
the tachyon will not result in any final state other than a single
supersymmetric type HO theory in ten dimensions.  
This supercritical orbifold does have other possible end-states. 
For instance, it can fragment into disconnected baby 
universes, of which precisely one supports supersymmetric type HO theory, with the others
containing unstable type HO$^{/}$ in various dimensions.  
However, such alternate endpoints 
are always fine-tuned.  Untuned transitions
lead to a unique and universal final state: a single component, supporting 
supersymmetric type HO in ten dimensions.
A phase space of absolute-stable transitions is depicted in Fig.~\ref{absolute}.

\begin{figure}[htb]
\begin{center}
\includegraphics[width=3.2in,height=2.5in,angle=0]{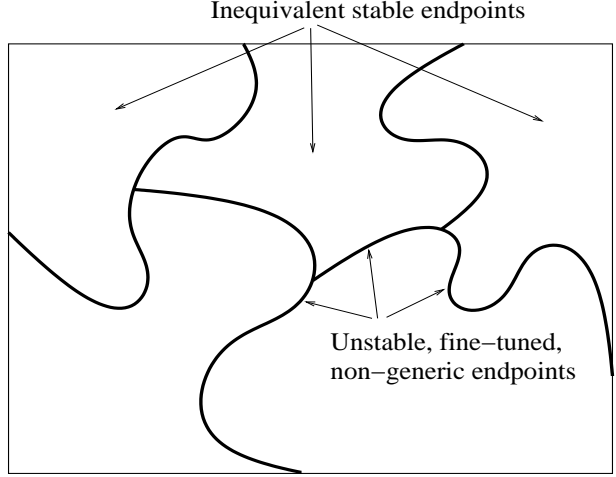}
\caption{Phase diagram of endpoints in locally stable transitions.  Contours depict loci
of non-generic, fine-tuned unstable endpoints of the transition.  The regions bounded by
these loci indicate stable but inequivalent endpoints.  For these purposes, the ``nothing'' state
will be considered a stable endpoint, in that nearby solutions converge to it at large $X\uu +$.   }
\label{local}
\end{center}
\end{figure}

We define a second type of stable transition as possessing 
the weaker property of being {\it locally stable} in
the space of solutions.  That is, a linearized perturbation around a locally stable
solution will always preserve the qualitative nature of the final state.  
However, a sufficiently
large deformation of a locally stable solution can lead to a qualitatively different 
but stable outcome.  
A model phase space of locally stable transitions is presented in Fig.~\ref{local}. 
The string theories of type HO$^{+}$ discussed in \cite{hellermanone} were shown to have the
property of local stability, but not absolute stability.  Starting with type HO$^{+}$ in
supercritical dimensions leads generically to a number of disconnected universes supporting
stable, supersymmetric type HO string theory.  However, the number of such universes,
and the chirality of the gravitini and gaugini in the baby universes, depends
on the number of zeroes of the tachyon configuration, as well as on
tachyon derivatives at the vanishing points.  
In some open sets of tachyon
configuration space there are no zeroes at all, and the universe is destroyed from within by 
a bubble of nothing.

\forceheading{Local stability of dimension-changing type UHE transitions}
We now wish to demonstrate that our dimension-reducing solutions of \uhe string theory
are locally stable, rather than absolutely stable transitions.
Consider a tachyon profile of the form $\ct(X) \equiv \m \, \exp{\b X\uu +} f(X\ll i, X\uu +)$, with
$i = 2,\cdots,9$, and $f(X\ll i, 0)$ taken to be a completely arbitrary real function.  We
assume that $:\ct(X):$ is an operator of definite weight $(\hh,\hh)$, which means that
$f(X\ll i, X\uu +)$ satisfies the diffusion equation
\bbb
\pp_+ f = \frac{\b\apr}{2} \pp_i^2 f \ .
\een{diffusion}
For $\m ~\exp{\b X\uu +} \gg 1$, the background is described by a string theory in
nine dimensions, supported on the vanishing locus of $f$.  For $ \sqrt{\apr} \gg |\pp_i f|$,
higher orders of conformal perturbation theory can be neglected, and the
leading order, given by equation \rr{diffusion}, is a good approximation to
the dynamics.  Using equation \rr{diffusion}, we demonstrate that two initial 
tachyon profiles can lead
to different outcomes for the final behavior of the system.

Consider one initial profile, given by $f(X\ll i, 0) = X\ll 8\sqd + X\ll 9\sqd - R\ll{\rm init}\sqd$, where we
assume $R\ll{\rm init}\sqd \gg \apr$.  Given this initial condition, the subsequent evolution of
the tachyon profile is
\bbb
f(X\ll i, X\uu +) = 
{1\over{\apr}}\left(
X\ll 8\sqd + X\ll 9 \sqd - R\ll{\rm init}\sqd + 2  \b\apr X\uu + \right) \ .
\eee
At a particular lightcone time 
$X\uu +\ll{\rm crit} =  {{R\ll{\rm init}\sqd} / ({2 \b\apr}})$,
the circle (defined by $\ct = 0$) shrinks to zero size, and the minimum of the
classical worldsheet potential rises above zero:
\bbb
\vws 	&=& \ct\sqd = \m\sqd \exp{2\b X\uu + } f\sqd 
\xxx
	&=& {{\m\sqd}\over{\apr\sqd}} \exp{2\b X\uu + } \lsqq
	X\ll 8\sqd + X\ll 9 \sqd +  2\b\apr (X\uu + - X\uu + \ll{\rm crit}) \rsqq \sqd\ ,
\eee
which is greater than $4\b\sqd \m\sqd (X\uu + - X\uu + \ll{\rm crit})\sqd $ when $X\uu + > X\uu +
\ll{\rm crit}$.
A positive-definite worldsheet potential increasing exponentially with
lightcone time has qualitative behavior akin to the bubble of nothing
described previously.  It does not describe the static $(8+1)$-dimensional 
heterotic string theory
at large $X\uu +$, or anything in the universality class thereof.

For a second initial profile, one may change the sign of the $X\ll 9\sqd$ term from $+1$
to $-a$, with $a>1$.  The tachyon profile then evolves as:
\bbb
f(X\ll i, X\uu + ) = f(X\ll i, 0) = {1\over{\apr}} \lsqq X\ll 8\sqd 
- a X\ll 9 \sqd  - R\sqd \ll{\rm init} + (1-a) \b\apr X\uu +\rsqq \ .
\eee
The effective value of $R\ll{\rm init}\sqd$ increases as $X\uu +\to \infty$, so the qualitative
behavior of the background at late times is that of the stable nine-dimensional heterotic
theory propagating in two disconnected baby universes, described by the two branches of
the hyperbola in the $X\ll{8,9}$ plane where $f(X\ll i, X\uu +)$ vanishes.

This establishes that two simple choices of initial conditions in the 
same linear dilaton background of the same theory can lead to qualitatively different behaviors
at late times, each of which is stable under small perturbations.  In one case, 
we have a universe-destroying 
bubble of nothing.  In the second case we obtain a bubble of new vacuum 
in which the universe bifurcates
into two stable nine-dimensional components.  This indeterminate behavior, with
multiple possible stable endpoints following from the same initial starting point,
is suggestive of a randomly populated landscape of vacua.

\section{The stable \HEnine~string theory in nine dimensions}
\label{further}
We have established that a final state of our dimension-reducing solution
describes a nine-dimensional theory, which we refer to as the \HEnine~theory.  
In the transition, the fields $X\ll 9,~\psi\ll 9$ and the
neutral fermion $\lt\nota$ have decoupled from the worldsheet,
and the dilaton gradient and string-frame metric
have acquired renormalizations associated with one-loop quantum corrections to
the free worldsheet theory.  As in the cases studied in 
\cite{previous2,previous3,previous4,previous5}, the effect is to transfer the central charge
contributions of the decoupled worldsheet degrees of freedom into the strength of the
dilaton gradient, keeping the total central charge constant.

In this section we analyze directly the \HEnine~final state
of our solution.  The final vacuum has a flat string-frame
metric and linear dilaton with spacelike gradient.  Furthermore, the final
vacuum has no tachyon degree of freedom.   The lightest excitations
are the metric $\hat{G}$, an NS two-form $\hat B$, the dilaton $\hat{\Phi}$,
an $E_8$ gauge field $\bf{\hat A}$ and a fermion $\bf{\hat \L}$ transforming
as a 16-real-dimensional Majorana spinor of $SO(8,1)$, in an adjoint
representation ${\bf 248}$ of $E_8$.  All of these fields descend in obvious ways
from the degrees of freedom present in the initial 10D \uhe theory.

We have also seen that the final $(8+1)$-dimensional state has a dilaton gradient
that is spacelike, with norm-squared $\hat{V}\sqd =  {1\over{4\apr}}$.
We can choose a new set of coordinates $Y\uu \macm,~\macm = 0,\cdots,8$ that put
the final dilaton and metric into a simple form; 
the appropriate coordinate transformations can be expressed 
in terms of the original $X^\pm$ variables as
\be
Y_8 & = & -\sqrt{2\apr} q X^- + \frac{\b\sqrt{\apr}}{2} X^+ \ ,
\xxx
Y^0 & = & \frac{2}{\b\sqrt{\apr}} X^- \ ,
\xxx
Y\ll \macm & = & X\ll\macm\ , \qquad \macm = 2,\cdots,8\ . 
\ee
In the $Y\uu \macm$ system, the $\macm,\macn,\cdots$ indices are raised and lowered with
the hatted metric $\hat{G}\ll{\macm\macn}$, and the dilaton is given by
\bbb
\hat{\Phi} = \hat{\Phi}\ll 0 + \hat{V}\ll \macm Y\uu \macm \ ,
\qquad
\hat{V}\ll {0,2,3,\cdots,8} = 0 \ ,
\qquad
\hat{V}\ll 1 = \hat{q} \ ,
\eee
with
\bbb
\hat{q} = {1\over{2\sqrt{\apr}}}\ ,
\qquad
\hat{G} \ll{\macm \macn} = \eta\ll{\macm \macn} \ .
\eee

\heading{Partition function}
The computation of the partition function for the \HEnine~theory proceeds as for
the \uhe theory. 
The major difference is that the contributions from the boson $X\ll 9$ and the fermions
$\psi\ll 9,~ \lt\nota$ are absent.  The linear dilaton does not couple to the
torus, and has no effect on the path integral.  The fermions $\psi\ll 9,~ \lt\nota$
are neutral under $\lmgg$, and have the same transformations under $\mfw$.  The
path integral on the torus with spin structure $a,b$ is therefore the same as that in the
\uhe theory, with one less factor of $(4\pi\sqd\apr\t\ll 2)\uu{-\hh} |\eta(\t)|\uu{-2}$
in the boson path integral, one less factor of $\yyt a b \uu\hh$ in the left-moving fermion
path integral, and one less factor of $\lrdd Z\uu a{}\ll b\rrdd\uu\hh$ in the right-moving
fermion path integral.

The left-moving fermion path integrals in the \HEnine~theory are thus
\bbb
\ellnine\uu 0{}\ll 0 (\tb) &=& {1\over{32}} \yyt 0 0 \uu {{15}\over 2}
\lsqq \yyt 0 0 \uu 8 + 31 \yyt 0 1 \uu 8
+ 31 \yyt 1 0 \uu 8 \rsqq \ ,
\xxx
\ellnine\uu 0{}\ll 1 (\tb) &=& {1\over{32}} \yyt 0 1 \uu {{15}\over 2}
\lsqq \yyt 0 1 \uu 8 + 31 \yyt 0 0 \uu 8
+ 31 \yyt 1 0 \uu 8 \rsqq \ ,
\xxx
\ellnine\uu 1{}\ll 0  (\tb) &=& {1\over{32}} \yyt 1 0 \uu 
{{15}\over 2} \lsqq \yyt 1 0 \uu 8 + 31 \yyt 0 0 \uu 8
+ 31 \yyt 0 1 \uu 8 \rsqq \ ,
\xxx
\ellnine \uu 1{}\ll 1  (\tb) &=& 0 \ .
\eee
The full path integral for fermions and superghosts, with spin
structure $a,b$ is 
\bbb
\effnine\uu a {}\ll b  (\t,\tb) \equiv  \xxt a b  \uu {7\over 2} \ellnine\uu a{}\ll b \ .
\eee
These are then combined into projected partition functions in NS and R sectors:
\bbb
\inine \ll{{\rm NS\pm}} &\equiv& \hh \lrdd \effnine\uu 0{}\ll 0 \mp \effnine\uu 0{}\ll 1 \rrdd \ ,
\xxx
\inine \ll{{\rm R+}} &\equiv& \inine\ll{{\rm R-}} = \hh \effnine\uu 1{}\ll 0  \ .
\eee
The path integral for bosons and reparametrization ghosts is
\bbb
i V\ll 9  (4\pi\sqd\apr\t\ll 2)\uu{-{9\over 2}} |\eta(\t)|\uu{-14} \ ,
\eee
and the measure for moduli takes the usual form.  The total partition function is
then:
\bbb
\int\ll{\cal F} {{d\t d\tb}\over{16\pi\sqd\apr \t\ll 2\sqd}}
 (4\pi\sqd\apr\t\ll 2)\uu{-{7\over 2}} |\eta(\t)|\uu{-14}
  \lrdd \inine\ll {{\rm NS} +} - \inine\ll{{\rm R}\pm} \rrdd \ .
\een{equation58}
The functions $F9^a{}_b$, $I9_{\rm NS\pm}$ and $I9_{\rm R\pm}$ exhibit 
the same modular transformation properties as their ten-dimensional counterparts
(see, e.g., Eqns.~(\ref{modinv1},~\ref{modinv2})),
so the modular invariance of Eqn.~(\ref{equation58}) follows.
Defining mass partition functions for the \HEnine~theory in parallel with
those of the \uhe theory, we find
\bbb
\znine\ll{\rm mass}\uu{\rm NS}(\t) &=&  (q\qb)\uu{+{1\over{16}}} 
\lsqq 1\com 785   + 108\com 500 ~(q\qb)\uu \hh
+ O\left(  q\qb \right) ~\rsqq \ ,
\xxx
\znine \ll{\rm mass}\uu{\rm R}(\t) &=& 1\com 984 + 4\com 058\com 880~
 (q\qb)\uu{1 } +  O\left(  (q\qb)\sqd \right)    \ .
\eee

We now pause to emphasize several points regarding the spectrum.
As expected, the NS sector is tachyon-free.  
The lowest NS states gain an effective mass-squared 
of $\hat V\sqd = {1\over{4\apr}}$  from their
coupling to the background dilaton gradient \cite{seiberg}, as is usual in subcritical
string theory \cite{polyakov1,polyakov2}. 
The first NS mass level consists of ${{7 \cdot 8}\over 2} - 1 = 27$ 
graviton polarizations, ${{7 \cdot 6}\over 2}$ B-field
polarizations, and one dilaton, together
with $7\cdot 248 =$ 1,736 polarizations of the $E\ll 8$ gauge
field, for a total of 1,785 physical states.  These states would be massless
in a background with constant dilaton.

The lowest Ramond mass level is a massless Majorana spinor in the adjoint of $E\ll 8$.  
A Majorana spinor in 9D has $16$ degrees of freedom off shell, which reduces to eight 
upon imposing the on-shell conditions.
Multiplying by the dimension of the adjoint representation, we obtain $8\cdot 248 =$ 1,984 
physical states. Interestingly, the spin-$\hh$ fermion does {\it not} 
obtain a nonzero mass from its coupling to the dilaton gradient.  

The masslessness of the lowest state is an inevitable consequence of the effective field
theory.  There is only a single Majorana adjoint fermion at the lowest level.
A single Majorana fermion can have no mass term with itself
in $8k$ spatial and one time dimension: if $C$ is the charge-conjugation matrix acting
on spinors, then $C$ and $C\G\uu \macm$ are
both symmetric.  A Majorana spinor $\bf{\hat{\L}}$ obeys
$C\ll{\a\b} {\bf\bar{\hat{\L}}}\ll\b = \hat\L\ll\a$, so terms such as $M\, {\bf{\bar{\hat{\L}}}}
\bf{\hat{\L}}$ and ${{\bf{\bar{\hat{\L}}}}}\, (\pp \hskip-.09in /\, \hat \Phi ) {\bf{\hat{\L}}}$
vanish identically by Fermi statistics.  Therefore, the rescaling 
from string frame to a canonically normalized fermion cannot introduce couplings that 
give $\bf{\hat{\L}}$ a nonzero physical mass.
We summarize the field content and spectral properties in the low-lying energy levels
of both the ten-dimensional parent UHE theory and the \HEnine~final
state in Table~\ref{spectrumtable}.

\begin{table}
\begin{center}
\begin{tabular}{|c||c|c|c|c|}
\hline
theory		& sector & mass  	 & field content				& mult.   \\
\hline\hline
\uhe 	& NS & $m^2 = -2/\apr$	& $\ct$						& 1	  \\
	& NS & $m^2=0$		& $\Phi (1)+ G (35) + B(28) +  {\bf A}(1984)$	& 2048   \\
		& R  & $m^2=0$		& ${\bf \L}_+ (1984) + {\bf \L}_- (1984)$	& 3968   \\
\hline
\HEnine	& NS  & $m^2=+1/(4\apr)$	& $\hat\Phi(1)+\hat G(27)+\hat B (21)+{\bf \hat A}(1736)$ & 1785   \\
	& R   & $m^2=0$		& ${\bf \hat\L}(1984)  $				& 1984   \\
\hline
\end{tabular}
\label{spectrumtable}
\end{center}
\caption{Summary of field content and multiplicities in the lowest-lying mass levels of the 
ten-dimensional UHE string theory and its nine-dimensional endpoint, following the transition.
The tachyon is absent in the \HEnine~final state.  Fields 
in the massless NS sector acquire an effective mass $m^2=1/(4\apr)$ after the transition, 
due to the coupling to the dilaton background.
Furthermore, the massless R sector loses half of its particle content in the transition. 
Interestingly, the lightest state in the \HEnine~theory is a fermion rather than a boson.
(Hatted quantities are reserved for the nine-dimensional theory.)}
\end{table}

\begin{table}
\begin{center}
{\footnotesize
\begin{tabular}{|c||c|c|c|c|}
\hline
string state & $m\sqd\equiv - k\ll \macm k\uu \macm$   &  field name  & $\begin{array}{c}
								{\rm transversality}/ \\
							         {\rm Dirac\ equation} 
								\end{array}$& gauge invariance\\
\hline\hline
${e\ll{\macm\macn} \tilde{\a}\ll{-1}\uu \macm \psi\ll{-1/2 }\uu \macn  } \kket{k~;0}\ll{1}$ & 
	$ {1\over{4\apr}}$
	& $\hat{G}\ll{\macm\macn}$,$\hat B\ll{\macm\macn}$,$\hat{\Phi}$ & $\begin{array}{c}
							(k + i \hat{V})\uu \macm e\ll{\macm\macn} =    \\
							(k + i \hat{V})\uu \macn e\ll{\macm\macn} = 0
							\end{array}$ &
                                                        $\begin{array}{c}
							\Delta e\ll{\macm\macn} = \tilde{\xi}\ll \macm 
                                                         (k - i \hat{V})\ll \macn    \\
							+ (k - i \hat{V})\ll \macm \xi\ll \macn
							\end{array}$  \\
\hline
$e\ll \macm \psi\ll{-1/2}\uu \macm\kket{k~;0}\ll{g\ll L}$ &  $ {1\over{4\apr}}$  &
  ${\bf \hat A}\ll \macm$ & 
$(k + i \hat{V})\uu \macm e\ll{\macm} = 0$
 & $\Delta e\ll \macm = (k - i \hat{V})\ll \macm \xi$ \\
\hline
$\kket {k~;\hat{\a}}\ll{g\ll L \cdot \mfw}  $
 & 0 & ${\bf  \hat \Lambda}\ll{\hat{\a}} $  & $ 
\pp \hskip-.08in /  {\bf \hat \Lambda} = 0 $  & -  \\
\hline
\end{tabular}}
\label{tab1}
\end{center}
\caption{{The lowest-lying normalizable string modes in the
\HEnine~theory.  The canonically normalized modes
of the graviton, dilaton and B-field acquire an effective mass $\Delta m =  {1\over{2\sqrt{\a}}}$
from their coupling to the background dilaton gradient, as does the $E\ll 8$ gauge field.
The $E_8$ adjoint fermion, on the other hand, does not acquire a positive mass: the canonically
normalized field $\bf{\hat\L}$ obeys the \it massless \rm Dirac equation.}} 
\end{table}


\heading{Absence of supersymmetry} 
The \HEnine~theory has no unbroken spacetime supersymmetry.  One way to see this is to 
demonstrate the complete absence
of Bose-Fermi mass degeneracy.  The $m\sqd = 0$ adjoint fermions ${\bf\hat\L}$ are split
from the gauge field ${\bf \hat A}$ in the spectrum by an amount 
$\Delta m\sqd = {1\over{4\apr}}$; the canonically
normalized modes of ${\bf \hat A}$ 
have an effective mass equal to ${1\over{2\sqrt{\apr}}}$.  Furthermore, the
multiplicities of the gauge field and adjoint fermions do not agree: 
for each of the 248 gauge generators, there are
seven physical polarizations of the gauge field, but eight physical polarizations of 
${\bf \hat \L}$, which transforms as a Majorana spinor of SO(8,1). 
There are also no spin-$1/2$ or spin-$3/2$ particles that could serve as 
degenerate superpartners
of the dilaton $\hat{\Phi}$, NS two-form $\hat B$ or graviton $\hat{G}$.  Indeed, the
lightest spin-$3/2$ fields and gauge-singlet spin-$1/2$ fields enter at
$m\sqd =  {4\over{\apr}}$, which is four times as heavy as the
normalizable excitations of $\hat{\Phi},~\hat B$ and $\hat{G}$.

Despite the lack of spacetime supersymmetry, our theory is tachyon-free.
The tachyon vertex operator would have to be built from a
matter primary of weights $(\tilde{h}, \tilde{h}
-1/2)$, with $\tilde{h} < 1$ prior to momentum dressing.  
The only left-moving fields with 
$\tilde{h} < 1$ are the 31 current algebra fermions $\tilde{\l}\uu A$,
which have $\tilde{h} = 1/2$.
(All twisted operators in the current algebra have weight at least one.)  
Each of the 31 fermions $\tilde{\l}\uu A$ transforms nontrivially under some 
element of the left-moving gauge group $\lrdd \IZ\ll 2\uu 5\rrdd\ll L$, and
none can therefore enter a physical vertex operator unaccompanied by other
current algebra fermions.  Hence, there are no tachyons in the \HEnine~final state.

Finally, the background has no moduli.  The lightest field is the dilaton, whose
normalizable excitations have $m = {1\over{2\sqrt{\apr}}}$.  Even the constant
mode of the dilaton $\d \hat{\Phi} = {\rm const.}$~does not represent a modulus in
the spacelike linear dilaton background.   Shifting $\hat{\Phi}$ can be compensated by
a redefinition of the spatial coordinate $Y\ll 8$, so even this
degree of freedom is not truly a modulus; rather, it is pure gauge.

\section{Discussion and conclusions}
\label{conclusions}
\begin{figure}[htb]
\begin{center}
\includegraphics[width=3.2in,height=3.2in,angle=0]{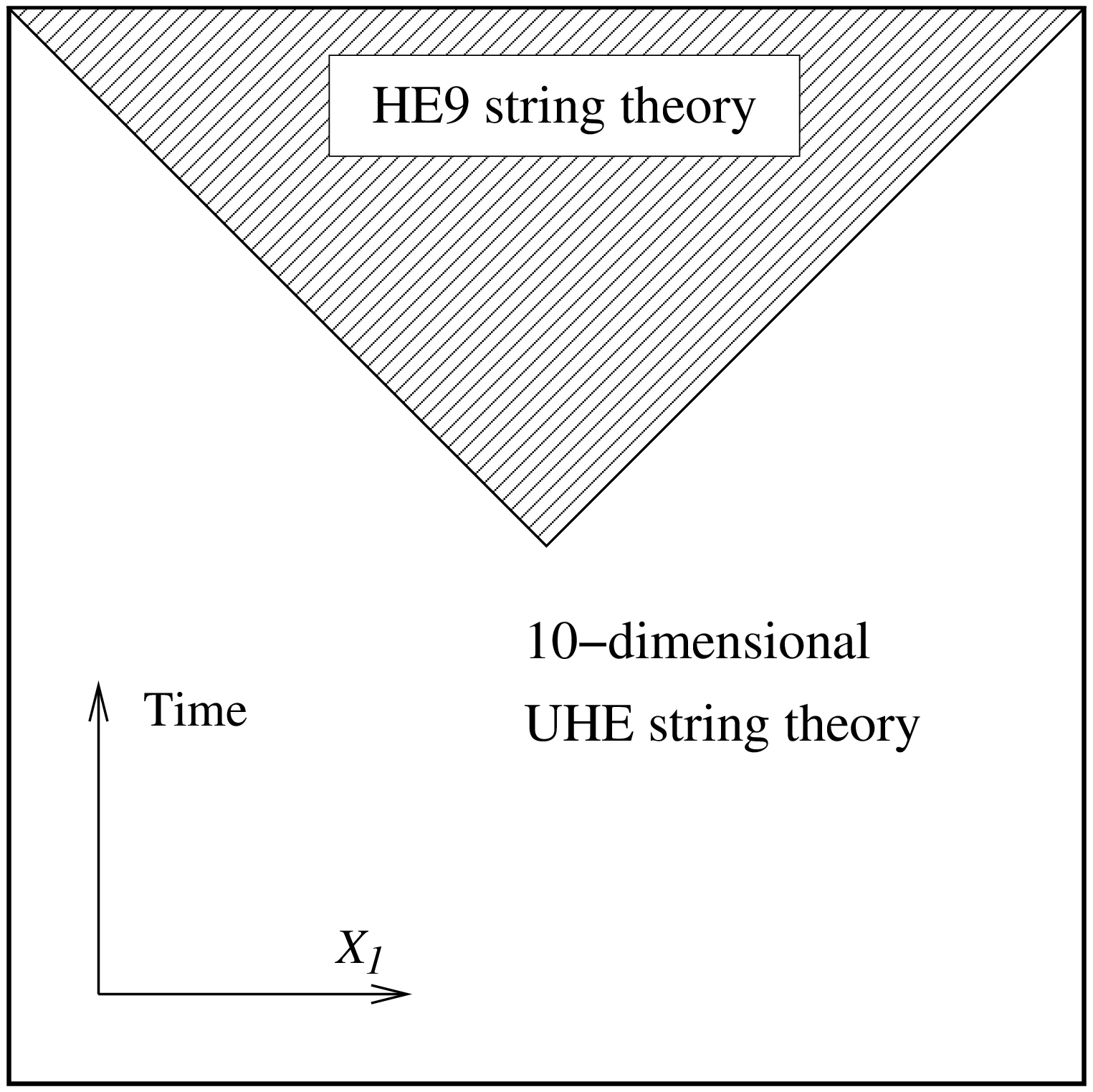}
\caption{The dynamical spacetime transition from ten dimensions outside 
a bubble wall to nine dimensions in the interior.  Our solution
focuses on the upper left-hand corner of the diagram, where the
bubble is a domain wall moving to the left at the speed of light.
The final phase is a stable theory with a single
$E_8$ gauge group and no spacetime supersymmetry.}
\label{bubble}
\end{center}
\end{figure}

In addition to the bubble of nothing solution studied by
Ho\v rava and Keeler, the \uhe theory in ten initial dimensions 
admits exact solutions that transit to a stable, nine-dimensional theory 
with no moduli and no spacetime supersymmetry.   This transition is depicted
schematically in Fig.~\ref{bubble}, where our solution focuses on the upper left-hand
region of the spacetime diagram.  

Generally, the absence of supersymmetry with no tachyons and no moduli
is interesting.  Few completely stable nonsupersymmetric
string theories are known in dimensions above $D=2$.
The $O(16)\times O(16)$ theory in ten dimensions is nonsupersymmetric
and tachyon-free at tree level, but its massless fields (such as the
dilaton and scale factor of the metric) acquire potentials from higher-genus string diagrams.
The \HEnine~theory, by contrast, has no moduli at tree level, and the effective mass shift 
due to the dilaton gradient renders the background stable against quantum corrections.
Background shifts due to higher-genus string
diagrams are finite away from the strong coupling region,
and can be incorporated via the Fischler-Susskind mechanism \cite{susskind1,susskind2}.

The qualitative nature of the final state of the \uhe string
depends on details of the initial tachyon profile.\footnote{As noted above,
similar non-deterministic behavior also occurs in the supercritical type HO$^{+}$ string
\cite{hellermanone}.}  This situation is reminiscent of the behavior
of an eternally inflating universe with a complicated scalar potential, in which
different, inequivalent vacua are populated by the random evolution of scalar fields across a
jagged landscape of local minima \cite{inflate1,inflate2}.  Tachyonic starting points
that can make locally stable (but not absolutely stable) transitions 
suggest a possible arena for studying the string landscape concept in a 
weakly-coupled limit.

Cosmological evolution in quantum gravity 
can produce qualitatively different outcomes from the same initial conditions. 
The stable \HEnine~theory is separated in phase space 
from the theory studied in \cite{horkeel,horkeel2}, though both theories descend from the
same unstable parent theory (the \uhe background).  
Two such basins of attraction must be separated by a transition with a characteristic
critical behavior.  
It would be interesting to find the set of unstable, fine-tuned endpoint 
theories that separate these
two stable attractors in phase space.  These theories lie along
critical loci in phase space, joining a patchwork of locally stable endpoints as
depicted in Fig.~\ref{local} above.

A broader question is,
what are the possible stable endpoints of
tachyon condensation descending from unstable parent theories other than type UHE?  
A census of unstable, ten-dimensional
heterotic string theories was taken by Kawai, Lewellen and Tye in 
Ref.~\cite{tye}.  They classified a total of six tachyonic backgrounds, 
with gauge groups $SO(32)$, $O(16)\times E_8$, $O(8)\times O(24)$, 
$(E_7 \times SU(2))^2$, $U(16)$ and $E_8$, the last belonging to 
the \uhe theory studied in this paper.  It would be interesting to analyze
the remaining unstable backgrounds in this classification, with the hope
of finding new string vacua of the type studied here.

Of the six, the \uhe theory is distinguished by having a sensible
11-dimensional interpretation \cite{horfab}.  In this interpretation, the
string-theoretic bubble of nothing is 
lifted to 11 dimensions as a cosmological spacetime with a particular geometry and
topology.  Given that the same \uhe state can also transition to the \HEnine~background,
the 11-dimensional description of type \uhe and its instabilities
could provide insight into noncritical and nonsupersymmetric
string vacua through the lens of M-theory.

\section*{Acknowledgments}
S.H.~is the D.~E.~Shaw \& Co.,~L.~P.~Member
at the Institute for Advanced Study.
I.S.~is the Marvin L.~Goldberger Member
at the Institute for Advanced Study.
The authors gratefully acknowledge additional support from 
U.S.~Department of Energy grant DE-FG02-90ER40542 (S.H.)
and 
U.S.~National Science Foundation grant PHY-0503584 (I.S.). 

\bibliographystyle{utcaps}
\bibliography{horava}


\end{document}